\documentclass[usenatbib]{mnras}
\usepackage{amsmath}
\usepackage{latexsym}
\usepackage{natbib,amssymb}
\usepackage{graphicx,todonotes}

\newcommand{\beq}{\begin{equation}\begin{aligned}}
\newcommand{\eeq}{\end{aligned}\end{equation}}

\newcommand{\angstrom}{\mbox{\normalfont\AA}}

\title[Wavelength Dependent PSFs]{Wavelength Dependent PSFs and their impact on Weak Lensing Measurements}

\author[Carlsten et~al.]{S. G. Carlsten$^1$\thanks{\tt scottgc@astro.princeton.edu},
Michael A. Strauss$^1$,
Robert H. Lupton$^1$,
Joshua E. Meyers$^{2,1}$,
\newauthor
Satoshi Miyazaki$^{3,4}$ 
\\$^1$Department of Astrophysical Sciences, 4 Ivy Lane, Princeton University, Princeton,
  NJ 08544
\\$^2$Kavli Institute for Particle Astrophysics and Cosmology, Department of Physics, Stanford
University, Stanford, CA 94305
\\$^3$National Astronomical Observatory of Japan, 2-21-1 Osawa, Mitaka, Tokyo 181-8588, Japan
\\$^4$Department of Astronomical Science, SOKENDAI (The Graduate University for Advanced Studies), 2-21-1
Osawa, Mitaka, \\Tokyo 181-8588, Japan
}

\begin{document}
\maketitle
\begin{abstract}

 We measure and model the wavelength dependence of the PSF in the Hyper Suprime-Cam (HSC) Subaru Strategic Program (SSP) survey. We find that PSF chromaticity is present in that redder stars appear smaller than bluer stars in the $g, r,$ and $i$-bands at the 1-2 per cent level and in the $z$ and $y$-bands at the 0.1-0.2 per cent level. From the color dependence of the PSF, we fit a model between the monochromatic PSF trace radius, $R$, and wavelength of the form $R(\lambda)\propto \lambda^{b}$. We find values of $b$ between -0.2 and -0.5, depending on the epoch and filter. This is consistent with the expectations of a turbulent atmosphere with an outer scale length of $\sim 10-100$ m, indicating that the atmosphere is dominating the chromaticity. We find evidence in the best seeing data that the optical system and detector also contribute some wavelength dependence. \citet{meyers15} showed that $b$ must be measured to an accuracy of $\sim 0.02$ not to dominate the systematic error budget of the Large Synoptic Survey Telescope (LSST) weak lensing (WL) survey. Using simple image simulations, we find that $b$ can be inferred with this accuracy in the $r$ and $i$-bands for all positions in the LSST field of view, assuming a stellar density of 1 star arcmin$^{-2}$ and that the optical PSF can be accurately modeled. Therefore, it is possible to correct for most, if not all, of the bias that the wavelength-dependent PSF will introduce into an LSST-like WL survey.

\end{abstract}
\begin{keywords}
gravitational lensing: weak --- atmospheric effects --- instrumentation: detectors --- methods: observational --- cosmology: observational
\end{keywords}

\section{Introduction} \label{sec:intro}

To make use of the statistical power of upcoming wide-field imaging surveys, systematic biases in cosmological weak lensing (WL) measurements need to be very well constrained. One such effect that can bias the WL measurement is the chromaticity of the point spread function (PSF), in the sense that the PSF changes in size and shape as a function of wavelength across a given broadband. Since the stars that the PSF is typically measured from have different SEDs from the galaxies that the cosmic shear is measured from, there is error in applying the stellar PSF to the galaxies when using broad-band data. Stellar and galactic PSFs can differ in size at the 1 per cent level for certain conditions (see below) which is well above the systematic error budgets for upcoming WL imaging surveys \cite[e.g.,][]{massey13}.  \citet{huterer2006} and \citet{amara2008} estimate that for a weak lensing survey with the coverage of the Large Synoptic Survey Telescope (LSST) \citep{lsst2009}, the average, uncorrected multiplicative bias to the shear needs to be $\lesssim0.003$ for the uncertainty in the estimates of cosmological parameters to be degraded by less than $\sqrt{2}$ times the purely statistical uncertainty. This corresponds to requiring that the systematic error in the PSF model size be constrained to $<0.1$ per cent \citep{paulin2008}.\par
	 \citet{cypriano10} first discussed the issue of PSF chromaticity and explored how the effective PSF for a galaxy was different than that for a star in a diffraction-limited telescope. \citet{meyers15} extended this work, using a more realistic model for the chromaticity of the PSF for ground based surveys. They use the standard Kolmogorov turbulent atmosphere result that the seeing scales as $\lambda^{-0.2}$ \citep{roddier81}. They find that the bias caused by using a stellar PSF can be corrected for at roughly the level of the LSST systematic error budget by correcting the PSFs on an `object-by-object' basis using multi-band photometry of each object and comparing with a large library of galactic SEDs. \citet{eriksen17} do a similar investigation assuming the filters of the \textit{Euclid} space-based survey \citep{euclid} and an optical model of \textit{Euclid} in which the size of the PSF $\propto\lambda^{0.55}$. \citet{voigt12} and \citet{semboloni13} consider how a wavelength-dependent PSF will couple with a galaxy's color gradient to bias the measurement of shape. \citet{plazas12} and \citet{meyers15} also discuss how differential chromatic refraction can bias shear measurements from the LSST by introducing an SED-dependent elongation of the PSF along the elevation vector. \par 
	The previous works all assume an \textit{a priori} model for the PSF as a function of wavelength. In this paper, we extend these previous studies by measuring the chromaticity of the PSF from ground-based data from the Hyper Suprime-Cam (HSC) survey. This can be done from stars of different color in the image frames. This is most important for ground-based surveys in which the atmosphere, optics, and detector together create the wavelength dependence, making this dependence nearly impossible to know \textit{a priori}.  \par
	
	The main concern is that the chromaticity due to the atmosphere will vary from night to night and possibly even within a night. The Kolmogorov atmosphere theory assumes a scale-free spectrum of turbulence but over-predicts the observed turbulence at larger scales \citep{tokovinin07, linfield01, conan00,  coulman88, ziad00}. The `von K\'{a}rm\'{a}n' turbulence model \citep{vonkarman} imposes an outer scale beyond which the wavefront structure function flattens and is a better fit to observational data \citep[e.g.,][]{oya16, book16}, as we demonstrate in this paper. The von K\'{a}rm\'{a}n model is found to adequately fit the PSFs present in the Sloan Digital Sky Survey imaging (Xin et al. in prep). The outer scale can vary from $\sim 10$ to $100$ m for common observatory sites \cite[e.g.,][]{ono17, conan00} and will change from day to day and even within a night \citep{linfield01}. This outer scale effectively steepens the chromaticity of the PSF over the Kolmogorov case \citep{martinez10, tokovinin02}. In Appendix \ref{app:chromo}, we show this along with a review of atmospheric turbulence as it applies to observed PSFs. \par
	
	In addition to the atmosphere, there is the possibility of chromaticity from the telescope and instrument, either from the optical design or construction or alignment errors. Diffraction-limited optics will have PSF size $\propto \lambda$. The charge diffusion present in CCDs will also contribute some chromaticity \citep{meyers2015b} due to the wavelength dependence of the photon absorption length in silicon. The relative contributions of the atmospheric chromaticity and that of the optics/detector will depend on the seeing, which is a function of time. \par

	In WL, both the ellipticity and overall size of the PSF need to be accurately modeled. Mis-modeling the PSF size will directly lead to a multiplicative bias in the shear. Mis-modeling the PSF ellipticity can result in an additive bias whose amplitude depends on the spatial correlations of the ellipticity errors \citep[for a recent review of WL systematics, see][]{mandelbaumreview}. In this paper, we consider only the simpler case of modeling the wavelength dependence of the PSF size. \par

 	There are other PSF effects that also need to be corrected for at this level of precision, including the brighter-fatter effect and variations of the PSF across the focal plane due to detector non-uniformities, optics, and atmospheric turbulence. The brighter-fatter effect comes from the lateral electric field generated from charge build-up in the pixel wells of the detector \citep{antilogus2014} and can be modeled by considering the correlations between pixels in flat-field images \citep{coulton17}. Spatial variations in the PSF are usually modeled as a low-degree polynomial function of position on a chip \citep[e.g.,][]{lupton2001, jee11} for each chip on the focal plane. Since the colors of stars should be independent of position on the sky, this effect is separable from the chromatic effect we investigate here. \par
	
	In Section \ref{sec:hscdata}, we analyze and discuss the chromaticity present in data from the Hyper Suprime-Cam Strategic Survey Program (HSC SSP). In Section \ref{sec:phosim}, we use simulations to address how well the chromaticity can be measured for various survey parameters. In Section \ref{sec:summary} we discuss our results and conclude.\par

\section{HSC Data} \label{sec:hscdata}
	The HSC SSP survey is an ongoing optical imaging survey in five broadbands ($grizy$) \citep{aihara17} with the Hyper-Suprime Camera \citep{miyazaki17, komiyama18, furusawa18, kawanamoto} on the 8.2m Subaru telescope operated by the National Astronomical Observatory of Japan. Because of its coverage, depth, and image quality (0.6 arcsec median seeing in $i$ band) the survey acts as a testbed and precursor for the upcoming Large Synoptic Survey Telescope (LSST) \citep{lsst2009}. In brief, the HSC instrument includes a Wide Field Corrector (WFC) which has an atmospheric dispersion corrector and delivers images with $<0.2$ arcsec FWHM (instrumental contribution only) across a 1.5$^\circ$ diameter field of view. The focal plane is paved with 104 2k $\times$ 4k science CCDs over 1.7 deg$^2$. The CCDs are 200$\mu$m thick Hamamatsu devices with 15$\mu$m wide pixels \citep{miyazaki17} subtending 0.168 arcsec. The survey consists of three layers with different sky coverages and depth: the wide layer will cover 1400 deg$^2$ and go to $\sim$ 26 mag in $r$, the deep layer will cover 26 deg$^2$ and go to $\sim$ 27 mag in $r$, and the ultra-deep layer will cover 3.5 deg$^2$ and  go to $\sim$ 28 mag in $r$. The first data release of the survey, consisting of data taken in 2014 and 2015, has been released to the public\footnote{The data release website is https://hsc-release.mtk.nao.ac.jp/} \citep{surveyrelease}. Details of the HSC data reduction pipeline can be found in \cite{bosch17}. The pipeline makes use of much of the software being developed for the LSST \citep{axelrod2010, juric2015, ivezic2008}. \par

	To investigate the level of chromaticity present in the HSC survey data, we measure PSF size as a function of color. To this end, we use single epoch, single band exposures (called `visits' in the pipeline). Roughly 500 visits in each of the bands are randomly selected from the S15B internal data release, which includes data taken over many nights in 2014 and 2015 from the wide, deep, and ultra-deep layers. The visits from the ultra-deep, deep, and about 1/3 of the wide-layer visits are included in the 2017 public data release \citep{surveyrelease}. Exposure times for individual visits in the wide layer are 150 s for the $g$ and $r$ bands and 200 s for the $i$, $z$, and $y$ bands. Exposure times for the deep layer are 180 s for the $g$ and $r$ bands and 270 s for the $i$, $z$, and $y$ bands. All bands have 300 s exposure times in the ultra-deep layer visits. Visits from each of the three layers are treated here identically. 
The seeing in the visits ranges from $\sim0.5$ to $\sim1.2$ arcsec. The HSC pipeline \citep{bosch17} identifies candidate stars for PSF modeling using a $k$-means clustering algorithm on the size of detected sources. Candidates are further restricted to sources brighter than 12,500 counts ($\sim$22.3 mag in $i$ in the wide layer). For this analysis, we just consider these candidate stars for PSF modeling. \par
	
\begin{figure*}
\includegraphics[scale=0.75]{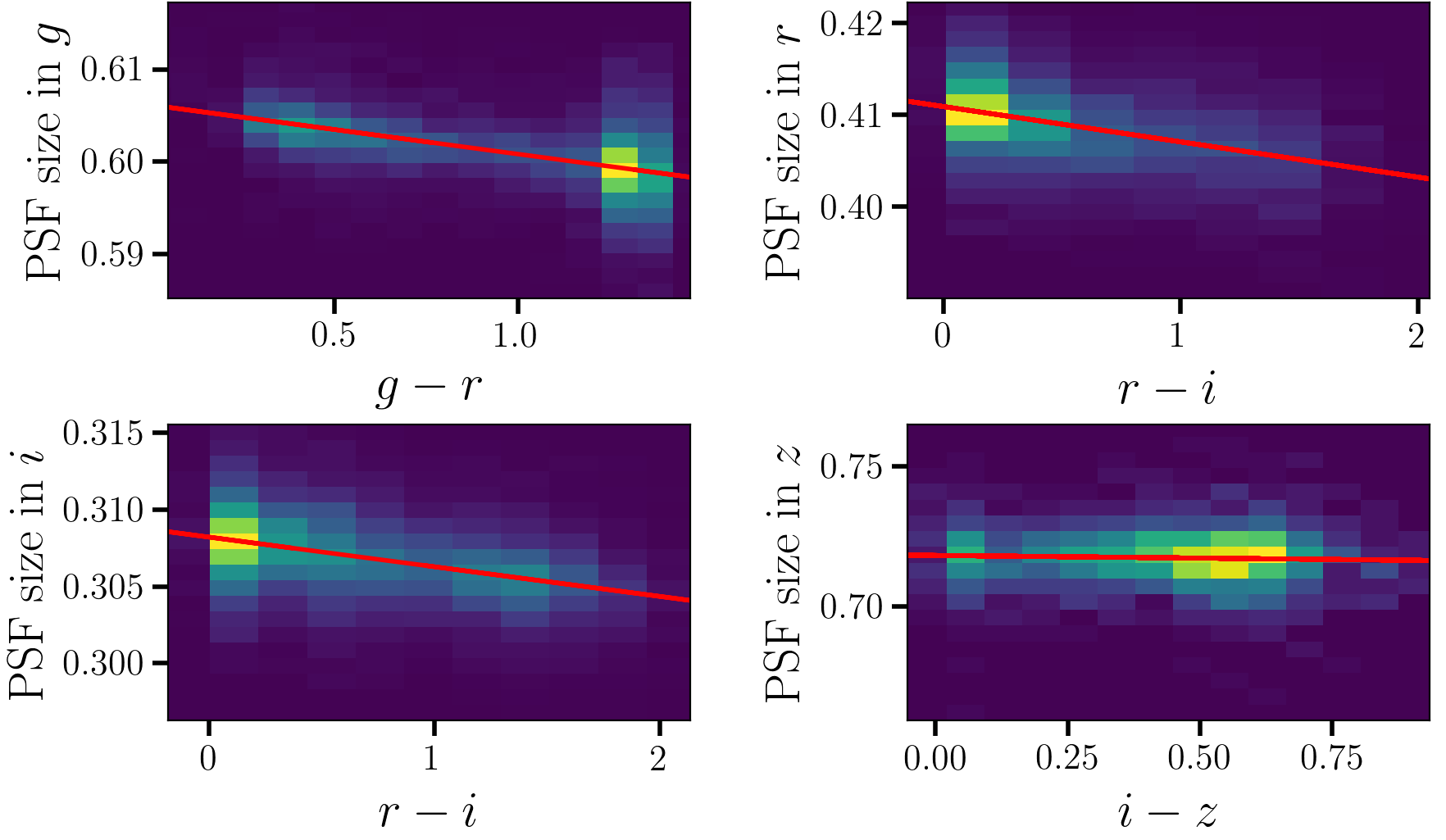}
\caption{2D histograms showing the source size (in arcseconds) versus color for stellar sources in one ($\sim1.7$deg$^2$) HSC visit in each of the four bands $griz$. Here the PSF size is quantified as trace radius measured from the weighted second moments of the image of each star rather than FWHM. FWHM is roughly $\sim1.5R$. The source size has been corrected for the spatial variation of the PSF across the HSC field of view. The red line is a linear fit to the data, highlighting the downward trend in points in the $g, r, $ and $i$-bands. The bimodal distribution of stars in $g-r$ is astrophysical and can be seen, for example, in the SDSS color-color diagrams of \citet{covey07}.}
\label{fig:hsc_dat_sample}
\end{figure*}	
	
	In the pipeline, the multiple visits in each band covering a given area are warped to a common pixel grid and coadded together. Object detection is run on this coadd and a catalog for the coadd is made. To get colors for each source, we cross-reference the single-visit catalog with the coadd catalog for that area. We expect that the PSF size in a band will depend on the slope of the source's SED across that band, therefore we use the color for each band that most closely represents that slope. For $g$-band visits we consider the $g-r$ color, for $r$ and $i$-band visits we consider the $r-i$ color, for $z$-band visits we consider the $i-z$ color, and, finally, for $y$-band visits we consider the $z-y$ color. We therefore further restrict the single-visit catalog to sources that appear in both coadd catalogs that make up the appropriate color. This leaves us with approximately 2000-8000 sources for each visit. Considering the 1.7 deg$^2$ field of view of HSC, this is approximately 1 source arcmin$^{-2}$.\par

	The HSC pipeline models the PSF with the PSFEx \citep{bertin13} software. Before this, however, the pipeline corrects for the brighter-fatter effect by using the pixel covariances found in flat-fields to reapportion the flux, as described in \citet{coulton17}. The PSF is modeled on a pixel basis and spatial variations of the PSF in a given visit are captured with a third degree polynomial function of the position on a chip. Sources of all color are used in the PSF modeling and so chromatic effects are not accounted for. The pipeline quantifies the size of a source with its `trace radius' defined as:
\beq
R \equiv \sqrt{I_{xx}+I_{yy}}
\eeq
where $I_{xx}$ and $I_{yy}$ are the two second moments. The second moments are measured with the Hirata-Seljak-Mandelbaum (HSM) adaptive weighting scheme of \citet{hirata2003} as implemented in GalSim \citep{rowe2015}\footnote{The calculated second moments are Gaussian weighted, so they deviate slightly from true second moments because the PSF is not described by a Gaussian. Nevertheless, these second moments are a robust measure of source size}. We use this definition for the PSF size throughout since it is directly relatable to the multiplicative bias in the shear caused by PSF mis-modeling \citep[e.g.][]{paulin2008}. For a Gaussian PSF, the FHWM is $\sim1.665$ times the trace radius. For a characteristic HSC PSF, the ratio is closer to FWHM $\approx1.5R$ due to the large non-Gaussian wings. \par

	The HSC pipeline provides size estimates of both the source and the PSF model at the location of the source. We use the PSF model to `correct' for the spatial variation of the PSF. To do this we subtract the PSF model size in quadrature from the measured source size for each PSF candidate star and then add the median PSF model size for the visit, using the whole 1.5$^{\circ}$ wide FOV. This should account for the spatial variations of the PSF without interfering with the chromatic variations, since those will appear in the residuals of the PSF model. \par
	
	 As shown in \citet{mandelbaum2017}, the pipeline PSF modeling is quite good, with modeling errors less than 2 per cent of the PSF size for almost all sources selected as PSF model candidate stars. The median PSF size error in the first year data is found to be $<0.4$ per cent in the $i$-band, which is within the requirements for first year weak lensing science. Figure \ref{fig:hsc_dat_sample} shows source size versus source color for one visit in each of the $g$, $r$, $i$, and $z$ bands. The $g, r,$ and $i$ band PSFs show a reasonably strong trend with color ($\sim$ 1 per cent across the color range), whereby redder sources are smaller. The $z$ and $y$ bands (the latter not shown) do not show much of a trend, for reasons that will be discussed below. \par

\begin{figure*}
\includegraphics[scale=0.62]{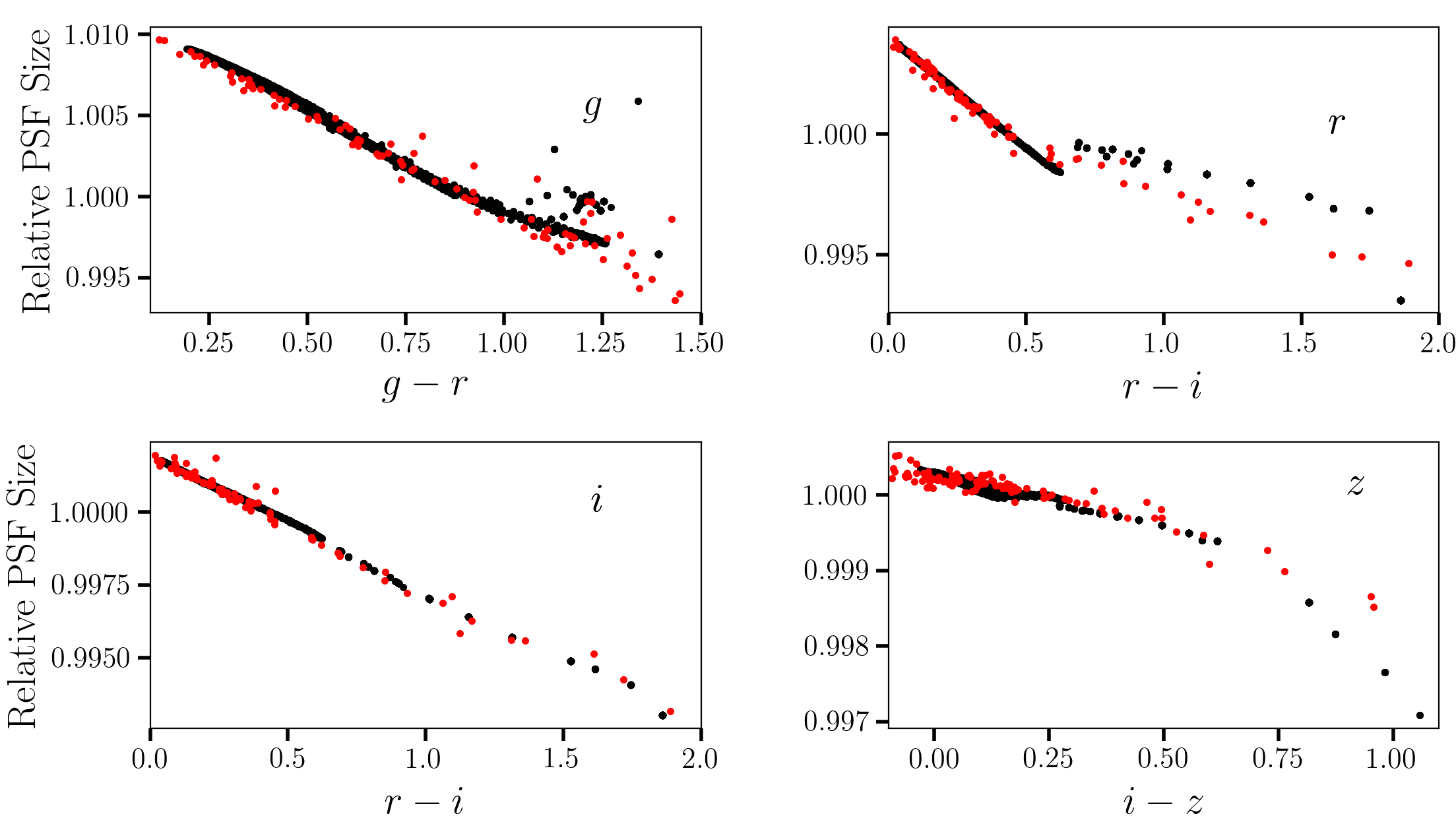}
\caption{Predicted PSF size (relative to the median) in the $griz$ bands versus color using the library of stellar SEDs described in the text. Black points are the synthetic library and red points show the Pickles library. The monochromatic PSF size is assumed to scale with wavelength as $R(\lambda)\propto \lambda^{-0.35}$. For the most part, the PSF size depends only on the color of the SED, and the slopes of the relation are the same in the synthetic and empirical SED libraries.}
\label{fig:sed_library}
\end{figure*}

\subsection{Model Fits}

	As stated in the introduction, if one can model the monochromatic PSF as a function of wavelength, then the chromatic bias in a WL survey can be removed by calculating `per-object' PSFs for objects with arbitrary SEDs. Therefore, we wish to infer the monochromatic PSF size vs. wavelength relations from the observed dependence of the broadband PSF size versus color. The observed PSF image in a band is the average of the monochromatic PSF weighted by the source SED. To convert between an observed size vs. color relation for stellar sources and a  monochromatic size vs. wavelength model, we need to use a library of stellar spectral energy distributions (SEDs). We use the library of the LSST Catalog Simulator\footnote{https://www.lsst.org/scientists/simulations/catsim} (CatSim) which is used in the LSST Photon Simulator (PhoSim) \citep{peterson15}. The SEDs are synthetic and use the \citet{kurucz1993} library down to an effective temperature of 4000K, and the low-mass models of \citet{baraffe2015} for cooler stars. The models include a range in metallicities from [Fe/H] $ = -4$ to [Fe/H] $ = 1$ for the Kurucz models, and range from [Fe/H] $=-4$ to [Fe/H] $=0.5$ for the low mass models. \par
	Since the relative populations of different types of stars (the number of high mass vs. low mass stars, for instance) might matter, we use the simulated universe of the LSST CatSim project to generate a realistic sample of SEDs. The Milky Way stellar population in CatSim uses the GalFast model of M. Juri\'{c} which is based on star counts using SDSS \citep{juric2008, ivezic2008} and has a realistic stellar color distributions. Samples of stars with 18$<i<$23 are generated in 0.1$^{\circ}$ radii circles at 10 random locations in the HSC survey footprint. This magnitude range roughly corresponds to the magnitude range of stars selected from the data. We use the combined catalog of model stellar SEDs of these samples in what follows. We find no significant difference in the following analysis if a different set of 10 locations are used or even if the empirical Pickles library \citep{pickles98} is used instead. \par

	We denote the monochromatic PSF size as a function of wavelength as $R(\lambda)$ and assume that it varies with wavelength in the form of a power law:
\beq
R(\lambda) \propto \left(\frac{\lambda}{5000\angstrom}\right)^{-b}
\label{eq:plaw}
\eeq
We denote the broadband PSF size as a function of color as $R(c)$. The observed PSF size of a star with SED $F$ and color $c$ in a band is\footnote{This expression ignores the fact that we use weighted second moments, but the error due to this approximation is only $\sim$1 per cent in the inferred wavelength dependence of the PSF.}:
\beq
R^2(c) = \frac{\int d\lambda G(\lambda) F(\lambda, c) R^2(\lambda)}{\int d\lambda G(\lambda) F(\lambda, c)}
\label{eq:size_v_color}
\eeq
where $G(\lambda)$ is the instrument's response in the band. Figure \ref{fig:sed_library} plots the PSF size from Equation \ref{eq:size_v_color} as a function of color for each SED in the library, assuming $R(\lambda)\propto \lambda^{-0.35}$ (which we find below is a reasonable model). Because there is close to a one-to-one correspondence between stellar color and SED, the PSF size is mostly just a function of the SED's color.\par

Nonetheless, to reduce the scatter seen in the SED library in Figure \ref{fig:sed_library} at the red edge, we consider only the restricted color intervals over which the PSF size-color relation is accurately linear and single-valued: $g-r\in$[0, 1.0] for $g$, $r-i\in$[0, 0.6] for $r$, $r-i\in$[0, 2.0] for $i$, $i-z\in$[0, 0.7] for $z$, and $z-y\in$[-0.1, 0.5] for $y$. These restricted ranges will make the analysis less sensitive to the relative populations of stars in the SED library. \par

We generate plots like Figure \ref{fig:sed_library} for several values of $b$ in Equation \ref{eq:plaw} and measure the slope in each case. The slopes are normalized by the $y$-intercept of the line to account for different normalizations of the PSF size (caused by different levels of seeing). This gives us a lookup table between the PSF size-color slope and the logarithmic slope $b$ in the PSF size-wavelength relation; the two quantities are almost linearly related. \par

This now gives us a way to infer the value $b$ for a given visit in a given band. We fit lines to the observed PSF size-color relation in each visit and in each band, as in Figure \ref{fig:hsc_dat_sample}. The slope of the PSF size-color relation is normalized by the $y$-intercept of the fitted line, and the index $b$ is inferred from the lookup table. The size-color relation is fit with a standard least squares minimization giving all sources equal weight, with two iterations of 3$\sigma$ clipping. Since there are unknown uncertainties from PSF modeling errors in addition to measurement uncertainties of the second moments, we cannot derive meaningful error bars from the fits directly. Instead, we address how accurately the wavelength dependence (i.e. the parameter $b$) can be determined using detailed simulations in \S\ref{sec:phosim}.\par

\begin{figure*}
\includegraphics[scale=0.65]{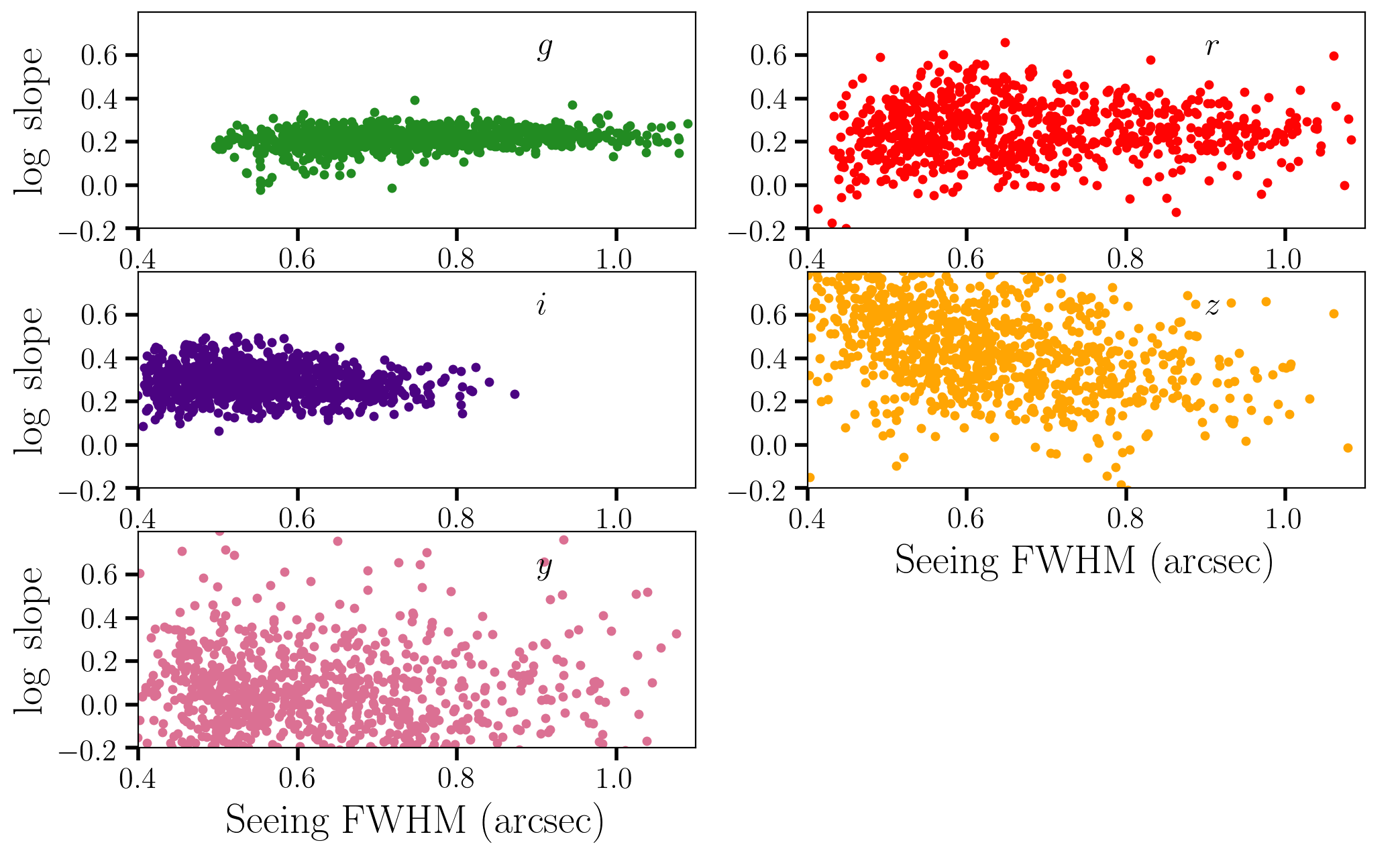}
\caption{The (negative) power law slope of equation \ref{eq:plaw} for all processed visits, i.e. the parameter $b$, as a function of the seeing in each visit. Each panel is a different band, as indicated in the corner of each subplot. Note that the $i$-band data have significantly better median seeing than the other bands \citep{aihara17}.}
\label{fig:hsc_slope_seeing}
\end{figure*}

In Figure \ref{fig:hsc_slope_seeing}, we plot the inferred power law slope, $b$, as a function of the seeing in each of the visits processed. To determine the seeing, the FWHM (not trace radius) is measured from the PSF model averaged over all of the chips for a particular visit. It is seen that the power law slope varies quite a bit between visits but is mostly in the range 0.0-0.5. The different bands show different trends with the seeing. In the $g$ and $r$ bands, there is a slight drop in the slope at very good seeing. In the $i$ and $y$ bands, there does not seem to be any trend. Finally, in the $z$-band, there is a very noticeable increase in the slope for good seeing. We will explore these trends in light of a simplified model of the optical/instrumental PSF in the next section.\par

	The scatter in the points also depends on the band, as well, being significantly greater for the $r$, $z$, and $y$-bands. There are several reasons for this: the color cut in $r$ significantly reduces the number of sources used, the $z$-filter is narrower than the other filters, there is more variation in atmospheric absorption in the $z$ and $y$ bands, and stellar SEDs all have similar slopes through the $z$ and $y$ bands which leads to a smaller difference in size between blue and red sources, making it difficult to measure the slope. While the PSF size of red and blue sources may differ by 1 per cent or more in the $g$, $r$, and $i$-bands, the difference is only of order 0.1 per cent in the $z$ and $y$-bands (Figures \ref{fig:hsc_dat_sample} and \ref{fig:sed_library}). In fact, many of the $y$-band visits in figure \ref{fig:hsc_slope_seeing} appear consistent with a power law slope of 0. Note that the range in seeing is much smaller in the $i$ band than the other four bands, due to the HSC observing survey strategy \citep{aihara17}; the $i$-band data are only taken in $<0.8$ arcsec seeing. \par

	The level of chromaticity appears to depend slightly on position in the focal plane. Considering either only sources in the inner 0.3$^{\circ}$ radius or the outer 0.2$^{\circ}$ of the field of view yields different average power law slopes only for the $r$ and $z$ bands (Table \ref{tab:inner_outer}); the difference is insignificant in $g$ and $i$. The $r$-band chromaticity is less steep in the outer regions, while the opposite is true for the $z$-band. We will explore why this might be in the next section.

\begin{table}
\caption{Median power law slopes, $b$, for all visits for sources within the inner 0.3$^{\circ}$, and the outer 0.2$^{\circ}$ of the field of view, respectively. The error is the standard error in the mean of the visits. The smaller number of sources in the $y$ band made many of the fits unstable, so it was not considered here. \label{tab:inner_outer}}

\begin{center}
\begin{tabular}{ccc}
\hline
Band & Inner 0.3$^{\circ}$ & Outer 0.2$^{\circ}$ \\
\hline
$g$ & 0.25 $\pm$ 0.003& 0.24 $\pm$ 0.002 \\
$r$ & 0.27 $\pm$ 0.008& 0.21 $\pm$ 0.006 \\
$i$ & 0.28 $\pm$ 0.004& 0.30 $\pm$ 0.003 \\
$z$ & 0.36 $\pm$ 0.01& 0.45 $\pm$ 0.01 \\
\end{tabular}
\end{center}
\end{table}

\subsection{HSC Optical Model}
	When the atmospheric seeing is very good, the instrumental contribution to the PSF will become more significant. In this section, we quantify this contribution in the context of the different behavior of the chromaticity as a function of seeing in different bands (cf. Figure \ref{fig:hsc_slope_seeing}). We will consider two sources of the instrumental contribution to the PSF: the image size produced by the wide-field corrector (WFC) and charge diffusion in the CCD detectors. The WFC gives an achromatic, but large, contribution to the instrumental PSF and thus dilutes other sources of chromaticity. Charge diffusion refers to the lateral spreading of the electrons as they move through the silicon layer. The amount of spreading depends on the distance traversed through the silicon. The expected diffusion has an rms spread of around 7$\mu$m, given the 200$\mu$m thickness of the HSC CCDs \citep{miyazaki17}. This spread is assumed to be independent of the brightness of a source (i.e. the brighter-fatter effect has already been corrected for as described above). Note that the HSC pixels are 15$\mu$m wide with pixel scale 0.17 arcsec per pixel so the diffusion is a small effect. Charge diffusion is a function of wavelength because photons of wavelength longer than 700nm penetrate a significant depth into the silicon before converting to an electron-hole pair. This leaves the electron with less remaining silicon to traverse, decreasing the charge diffusion. The amount of charge diffusion is smaller if the CCDs are run with a higher bias voltage but we assume that this does not change between visits. \citet{miyazaki17} performed tests on the constructed HSC system and provide estimates of the contribution of the WFC and charge diffusion to the optical PSF. In Tables \ref{tab:wfc} and \ref{tab:diffusion} we reproduce their Tables 4 and 9 describing the image size specifications of the WFC and amount of charge diffusion, respectively. To go from the 80 per cent encircled energy diameter of Table \ref{tab:wfc} to a FWHM, we assume the instrumental PSF is a Gaussian so $D_{80}$=1.5FWHM.\par
	
	It is worth noting that the wavelength-dependent absorption length of photons in silicon can lead to another, opposite chromatic effect. Because the HSC beam is fast (f/2), photons will be incident onto the silicon at a fairly steep angle. Since redder photons will be absorbed over a wider range of depth in the silicon, the lateral spread of photons in the incident beam can lead to a larger image size for longer wavelengths. This effect is essentially due to the fact that red photons will not all be absorbed at the depth of best focus within the silicon. \citet{meyers2015b} argue that for the LSST this effect will dominate over the charge diffusion effect due to the extremely fast f/1.2 LSST beam. For HSC however, \citet{miyazaki17} estimate that this effect's contribution to the total instrumental PSF is roughly a factor of 3 less than that from charge diffusion, at least at the telescope boresight. It is unclear whether this is still the case near the edge of the field of view where the incident angle of the photons will be steeper, but here we assume that it is and ignore this additional chromatic effect.

\begin{table}
\caption{Table 4 from \citet{miyazaki17}. Image size delivered by the WFC, not including the contribution from the atmosphere. Estimated sizes are given as D$_{80}$, the diameter enclosing 80\% of image flux and consider the `worst case' throughout the field of view. Sizes for telescope elevation of 90$^\circ$ and 30$^\circ$ are included.\label{tab:wfc}}

\begin{center}
\begin{tabular}{cccc}
\hline
Band & Central $\lambda$ [\angstrom] & EL=90$^\circ$ [arcsec] & EL=30$^\circ$ [arcsec] \\
\hline
$g$ & 4700 & 0.199 & 0.234\\
$r$ & 6200 & 0.197 & 0.232\\
$i$ & 7600 & 0.200 & 0.233\\
$z$ & 9100 & 0.206 & 0.214\\
$y$ & 10200 & 0.210 & 0.228\\
\end{tabular}
\end{center}
\end{table}

\begin{table}
\caption{Table 9 from \citet{miyazaki17}. Image Size due to charge diffusion in the CCDs.\label{tab:diffusion}}
\begin{center}
\begin{tabular}{ccc}
\hline
$\lambda$ [nm] &  rms spread [$\mu$m] &  FWHM [arcsec]\\
\hline
$700$ & 6.9 & 0.18\\
$800$ & 6.6 & 0.17\\
$900$ & 5.8 & 0.15\\
$1000$ & 4.5 & 0.12\\
\end{tabular}
\end{center}
\end{table}

\begin{figure}
\includegraphics[scale=0.35]{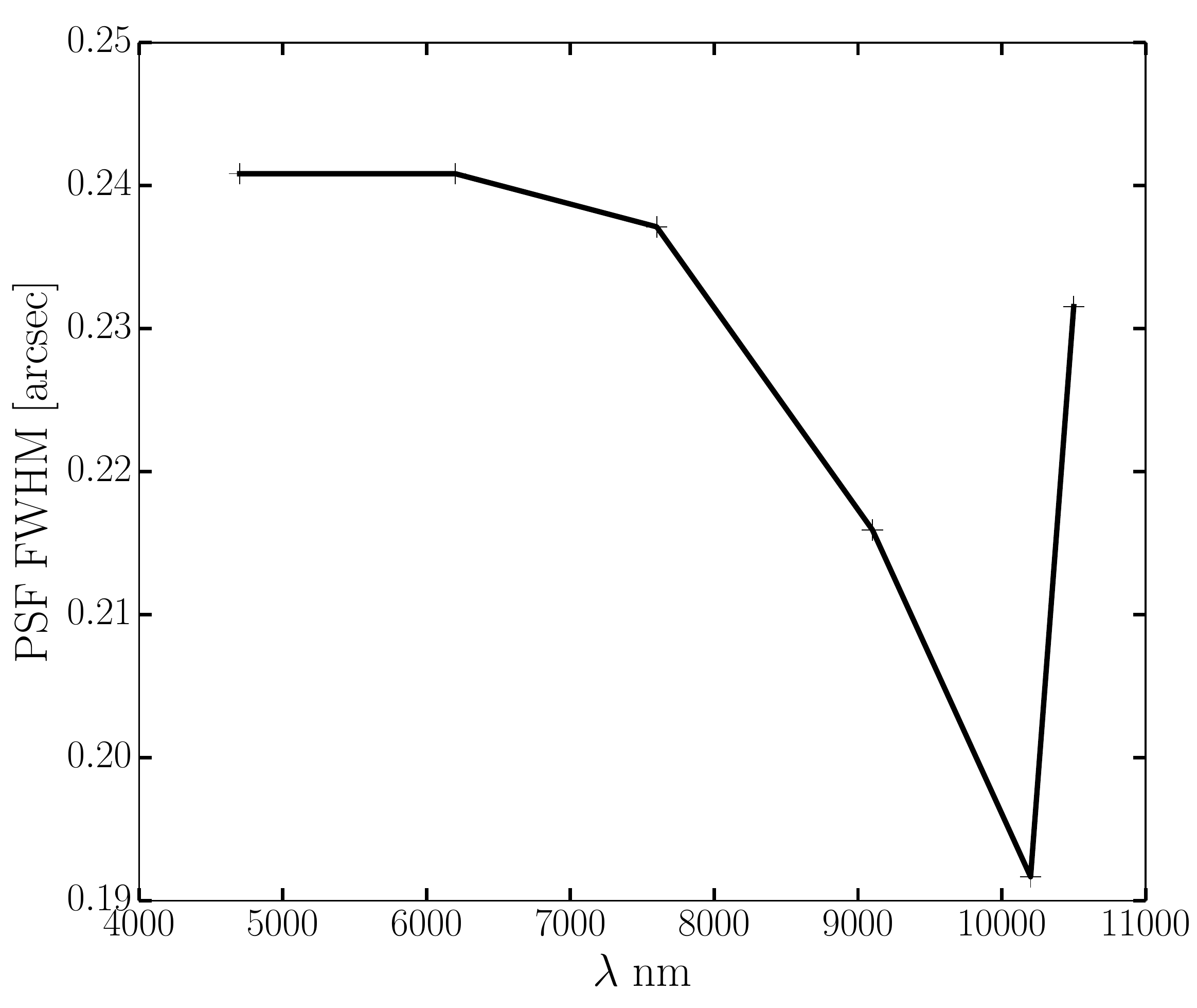}
\caption{The assumed width of the instrumental component of the PSF at different wavelengths. The spike at 10500$\angstrom$ represents the assumed increase in PSF size at the redward edge of the $y$-band (see text). }
\label{fig:instr_psf}
\end{figure}

Combining Tables \ref{tab:wfc} and \ref{tab:diffusion}, the final instrumental PSF widths used are shown in Figure \ref{fig:instr_psf}. We use image sizes for the WFC at a telescope elevation of 30$^\circ$. While this is conservative, we are not including other contributors to the PSF, such as tracking errors and wind jitter. To measure the slope of the PSF size across the whole $y$-band, we add another data point at 10,500$\angstrom$, extrapolated from the trend from $z$ to $y$. The nearly complete transparency of silicon to photons of this wavelength is expected to cause a significant increase in PSF size on the redward edge of $y$ as explained below. The final value we choose (see Figure \ref{fig:instr_psf}) is a little arbitrary but seems to fit the data. \par 

We can combine the instrumental PSF estimate along with predictions for a turbulent atmosphere to understand the trends seen in Figure \ref{fig:hsc_slope_seeing}. To do this, we generate atmospheric PSFs with a given outer scale length and with different levels of seeing as discussed in appendix \ref{app:chromo} and then convolve this PSF with a Gaussian representing the optical/instrumental components combined together in quadrature\footnote{A Gaussian accurately represents charge diffusion's contribution to the PSF, while the optical component will have more high frequency power. For this analysis, however, a Gaussian is likely to be an adequate approximation.}. Monochromatic PSFs for several different wavelengths are generated in each band for a range of seeing values. The trace radii of these PSFs are measured and a power law is fit as a function of wavelength. We can then plot the expected power law slope versus the average FWHM seeing in that band.  \par

\begin{figure}
\includegraphics[scale=0.66]{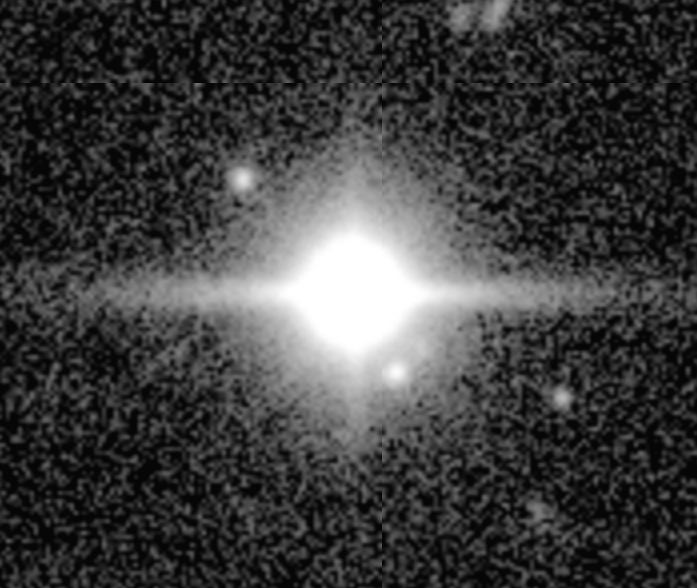}
\caption{An example of the `cross'-shaped pattern around bright sources in the $y$-band due to diffraction within the CCD. The large horizontal spike is parallel to the serial register of the CCD. The star is not saturated and is characteristic of a $\sim19^{th}$ mag star in $y$.}
\label{fig:yband}
\end{figure}

\begin{figure*}
\includegraphics[scale=0.63]{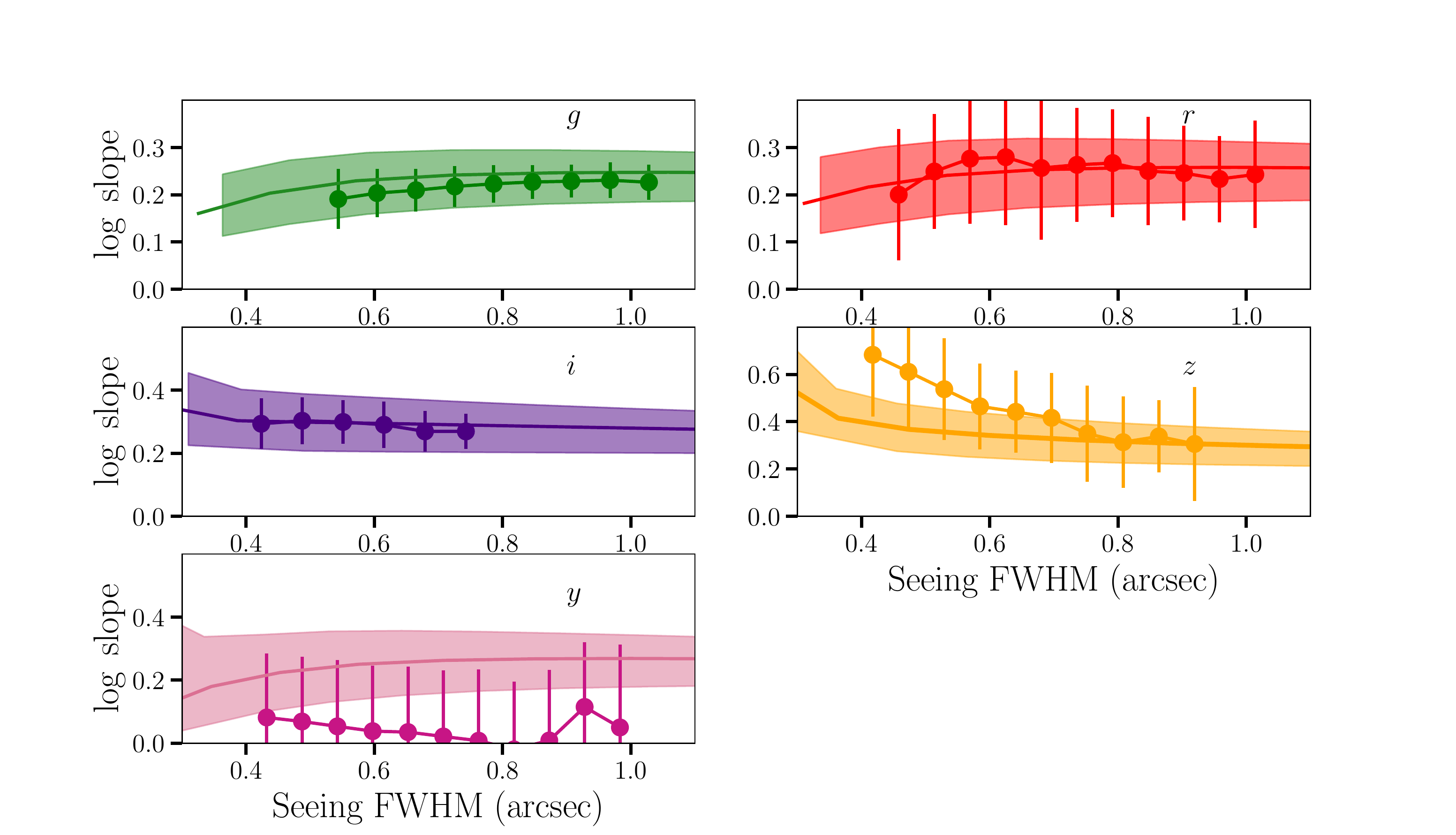}
\caption{The power law slope of the PSF size versus wavelength relation, $b$, as observed and as modeled. The data points are the data from figure \ref{fig:hsc_slope_seeing} smoothed by a running mean of box size 0.1 arcsec. The error bars represent the standard deviation in each bin. The model includes contributions from the atmosphere and the HSC instrument. The colored region gives the model predictions ranging from an outer scale length of 8m (top) and a Kolmogorov atmosphere (bottom) and the inner line is the prediction for an outer scale length of 25m. The model does a good job of fitting the trends seen in $g$, $r$, $i$, and, to a lesser extent, $z$. In $y$, the diffraction within the CCD seen in Figure \ref{fig:yband} obscures the chromatic effect.}
\label{fig:hsc_dat_th}
\end{figure*}

Figure \ref{fig:hsc_dat_th} shows the data of Figure \ref{fig:hsc_slope_seeing} smoothed with a running mean, along with the expectations for an atmospheric outer scale length of 8m, 25m, and the Kolmogorov result with no outer scale, all convolved with the instrumental PSF of Figure \ref{fig:instr_psf}. 25m is the characteristic outer scale length for many observatory sites, including Maunakea \citep{ziad00, ono17}. With the exception of $y$-band, the models seem to predict, at least qualitatively, the behavior in the different bands. The models predict a drop in the log slope at good seeing that is visible in the data in the $r$ and somewhat in the $g$ band. The instrumental PSF shows less chromaticity than the atmospheric PSF in these bands, so the level of chromaticity drops (smaller $b$) as the seeing gets better. \par

The opposite situation occurs in the $z$-band where the level of charge diffusion depends steeply on wavelength, as shown in Table \ref{tab:diffusion}. Here the instrumental PSF has a steeper wavelength dependence than the atmospheric PSF, leading to increased slope at better seeing. The same situation should be true in the $y$ band, but the increased transparency of the silicon to photons on the redward edge of $y$ is so great that these photons actually bounce off the bottom of the silicon and get absorbed on the way back to the top surface. These photons get diffracted off the gate pattern on the bottom of the silicon, leading to a large cross-shaped pattern containing roughly 2 per cent of the flux around stars in HSC $y$ images (J.E. Gunn, private communication). An example of this pattern is shown in Figure \ref{fig:yband}. This leads to a larger instrumental PSF on the red edge of $y$, as we have assumed in our model (cf. Figure \ref{fig:instr_psf}). The diffracted light increases the measured PSF trace radius in $y$ by 0.1-0.2 per cent, roughly canceling out the chromatic effect of the atmosphere and leading to little net chromaticity in this band. \par

	We can now partly understand the spatial variation of the wavelength dependence shown in Table \ref{tab:inner_outer}. Figure 4 of \citet{miyazaki17} shows the designed performance of the WFC as a function of position in the focal plane. $r$ and $z$ show a significant image size difference between the inner 0.3$^{\circ}$ and outer 0.2$^{\circ}$ regions, while $g$ and $i$ do not. Since the instrumental component of the PSF in the $r$ band is essentially wavelength-independent, the overall chromaticity decreases in the outer region of the focal plane as the instrumental PSF dilutes the chromaticity from the atmosphere. In the $z$ band, the instrumental PSF is steeply chromatic, but it is still dominated by charge diffusion which is constant throughout the focal plane. It is therefore unclear why the wavelength dependence is steeper in the outskirts of the focal plane in the $z$ band. \par

\section{Simulations} \label{sec:phosim}
	In the previous section, we showed that the PSF is significantly chromatic in the HSC-SSP survey. Other works \citep[e.g.,][]{meyers15, eriksen17} have shown that the wavelength dependence of the PSF can be corrected for in WL analyses if it is known. \citet{meyers15} argue that since a size-versus-wavelength relation of the form $R \propto \lambda^{-b}$ with $b=0.2$ leads to a systematic error, if uncorrected, that is roughly 10$\times$ greater than the systematic error budget of the LSST WL shear survey, the power law slope needs to be known to $\Delta b \sim 0.02$. A similar but even tighter constraint is found in \citet{eriksen17} for \textit{Euclid}. In the previous section we infer this power law slope from the dependence of the PSF size on color and get physically reasonable values. However, since we do not know the ground truth of the wavelength dependence, we cannot conclude whether the inferred slope is accurate enough to correct for the PSF chromaticity. To answer this question, we require simulations where the ground truth can be known. In this section we explore how many stars are needed to determine $b$ to $ \sim 0.02$. While this analysis is applicable to other imaging surveys, we focus on the LSST survey since chromatic biases are really only a significant concern for a survey with the coverage of LSST.\par

\subsection{Accuracy of the Inferred Wavelength Dependence}
	We do simple image simulations whereby we make a series of postage-stamp images of stars and measure the size and flux of these images after noise has been added. We start with a catalog of stellar SEDs with magnitudes within the range 17$<i<$22, generated in a 1$^{\circ}$ radius circle at ($\alpha=0.0^{\circ}, \; \delta=0.0^{\circ}$) using the CatSim Milky Way simulator. This magnitude range will be detectable in a single LSST visit (with S/N$\gtrsim$30) while not saturating. The simulator provides estimates of the AB magnitude at 5000$\angstrom$ for each star and from this we calculate the expected number of counts in each image using the collecting area of LSST, the exposure time (30 seconds) for each visit, and the gain (which we take to be 1.7e$^{-}$/ADU). Poisson noise is added to each pixel in the postage stamp along with noise associated with the sky foreground photons, estimated using the LSST Photon Simulator \citep[PhoSim;][]{peterson15}.  \par
 
As we have seen, the atmospheric PSF dominates the chromaticity, and so we include only this component in the simulations. Since we are here focused on the PSF size, we are not interested in the effects of short exposure time (which should affect PSF ellipticity more \citep[e.g.,][]{heymans12, chang13}) and we will consider the long-exposure limit PSFs. We assume a von K\'{a}rm\'{a}n turbulence power spectrum for an outer scale length of 10m (which is on the short side of what is to be expected). The atmospheric PSFs are generated by Fourier transforming the resulting atmospheric structure function. For each band, atmospheric PSFs are generated at several wavelengths. These monochromatic PSFs are combined with each stellar SED to create the broad-band PSF for each object. 
 The second moments are measured from the postage stamps using the same HSM algorithm as in the HSC data. We also add some scatter in the measured size due to errors in the PSF modeling. We consider both the cases where the PSF modeling is perfect and also when it is accurate to an rms of $\pm1$ per cent, which is characteristic of the PSF modeling in the HSC survey \citep{mandelbaum2017}. By fitting a power law to the size of the monochromatic PSFs, we can determine the `ground truth' chromaticity and compare to what we infer from the broad-band images of the stars. \par
 
 \begin{figure}
\includegraphics[scale=0.47]{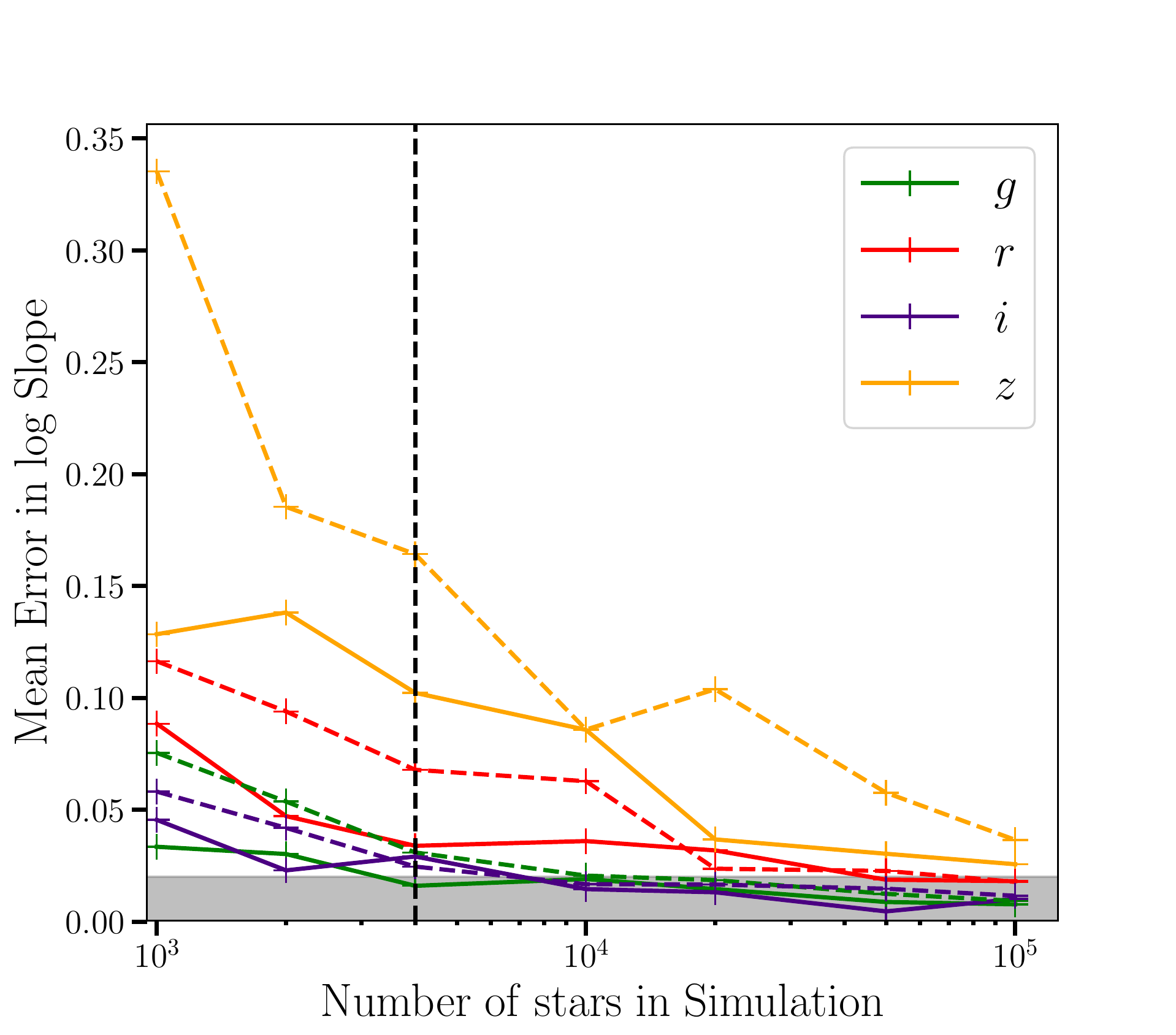}
\caption{The average of the absolute value of the error in the determination in the power law slope of the PSF size-wavelength relation as a function of the number of 17$<i<$22 stars used in the simulation. The final number of stars used in the analysis is somewhat less due to the flux and color cuts, as described in the text. The dashed lines are for the case where there is $\pm$1 per cent rms error in the modeling of the PSF and the solid lines are for perfect PSF modeling. An outer scale length of 10m is assumed for the atmosphere. The `ground truth' wavelength dependence for the simulations is $R\propto\lambda^{-0.31}$ for $g$, $R\propto\lambda^{-0.33}$ for $r$, $R\propto\lambda^{-0.35}$ for $i$, and $R\propto\lambda^{-0.36}$ for $z$. The gray area denotes the goal of $\Delta b \leq 0.02$. The vertical black line represents 1 star arcmin$^{-2}$ for an 0.3$^{\circ}$ wide annulus at half the radius of the LSST focal plane (see text).  The average error goes down as roughly $N_{\rm stars}^{-0.5}$.}
\label{fig:slope_err_sim}
\end{figure}

We can check how the accuracy of the inferred chromaticity depends on the number of stars simulated. As in the HSC pipeline, simulated stars are limited to those with counts $>10,000$ (S/N$\gtrsim60$) which reduces the number of stars by $\sim$25 per cent. The measured size of the stars is plotted against the measured color and the chromaticity is inferred in the same way as in Section \ref{sec:hscdata}, using the LSST filter response curves to generate the look-up table. Figure \ref{fig:slope_err_sim} shows the resulting average of the absolute value of the error in the inferred logarithmic slope of the size-wavelength relation (i.e. the parameter $b$) as a function of the number of 17$<i<$22 stars used in the simulation. The mean error in the logarithmic slope goes down if more stars are simulated, as expected. Considering that 2,000-8,000 stars were used in each HSC visit in Section \ref{sec:hscdata}, the power law slopes we derived in that section are likely to be accurate to $\Delta b \sim 0.07$ in $r$, $\Delta b \sim 0.02$ in $i$ and $g$, and $\Delta b \sim 0.1$ in $z$, assuming the modeling of the PSF is accurate to $\pm1$ per cent rms. These different levels of accuracy are due to a combination of the width of the band, the width of the color cut used, and the variation in the slope of the stellar SEDs across that band. This trend is consistent with the fact that the $r$ and $z$ bands show more scatter than $i$ and $g$ in the HSC data, as we saw in Figure \ref{fig:hsc_slope_seeing}. \par
 
 To understand what Figure \ref{fig:slope_err_sim} means for LSST, it is important to know how many stars are likely to be in each LSST visit. CatSim predicts $\sim$20,000-80,000 stars with $17<i<22$ in the whole 3.5$^{\circ}$ diameter LSST FOV for galactic longitudes $80^{\circ}>|b|>40^{\circ}$. These numbers correspond to roughly 0.5 to 2 stars arcmin$^{-2}$. If all these stars are used to determine the PSF-wavelength relation, it would be sufficient to determine $b$ to $\Delta b \sim 0.02$ in at least the $g$, $r$, and $i$ bands, even when PSF modeling errors are considered.

\subsection{Variations in the Chromaticity with Focal Plane Position}
	We found evidence for a dependence of PSF chromaticity on position in the focal plane of HSC. The same might be true for the LSST. To consider the optical component of the PSF, we use PhoSim to simulate the chromaticity at different positions in the focal plane. PhoSim forces the wavelength dependence of the PSF to have the Kolmogorov $\lambda^{-0.2}$ result, therefore we replace PhoSim's in-built atmospheric simulation with our own that produces a more realistic wavelength dependence. For a simulation, 500 phase screens are generated in the usual Fourier way \citep[e.g.,][]{mcglamery1972} with a von K\'{a}rm\'{a}n power spectrum for the central wavelength in that band. The phase screens are masked with the expected LSST pupil and Fourier transformed to give instantaneous atmospheric PSFs. The instantaneous PSFs of these 500 phase screens are ensemble averaged to give a long-exposure atmospheric PSF. Deflection angles are then drawn from this PSF and given to photons as they pass through the atmosphere. The wavelength dependence is included by scaling the deflection angle by a term proportional to the ratio of the PSF size at the photon's wavelength to that at the band's central wavelength. This ratio comes from the numerical fitting of \citet{tokovinin02}\footnote{This scaling implicitly assumes that the wavelength dependence is simply a dilation/contraction of the PSF. This is strictly true for the Kolmogorov PSF but is not true for the von K\'{a}rm\'{a}n PSFs. For the latter, redder wavelengths have smaller PSFs but their wings become more pronounced causing the trace radius to fall more slowly with wavelength than the FWHM. Therefore, using Equation \ref{eq:toko} will overestimate the chromaticity by 5-10 per cent in the parameter $b$. For the purpose of this simulation, this is irrelevant since we are not focused on the absolute level of chromaticity but on how the chromaticity changes with position in the focal plane. }:

\beq
\epsilon_{\rm vK} = \epsilon_{\rm Kolm} \sqrt{1-2.183\left(\frac{r_0}{L_0}\right)^{0.356}}
\label{eq:toko}
\eeq
where $\epsilon_{\rm vK}$ is the PSF FWHM at the desired wavelength, $\epsilon_{\rm Kolm}$ is the PSF FWHM for Kolmogorov turbulence ($\propto \lambda^{-0.2}$), $r_0$ is the usual Fried's parameter ($\propto \lambda^{1.2}$, see Appendix \ref{app:chromo}), and $L_0$ is the outer scale length. \citet{tokovinin02} finds that Equation \ref{eq:toko} is accurate to 1 per cent for most values of $r_0$ and $L_0$. This equation has been validated in more recent works \citep{martinez10, martinez2014}.\par

\begin{figure}
\centering
\includegraphics[scale=0.34]{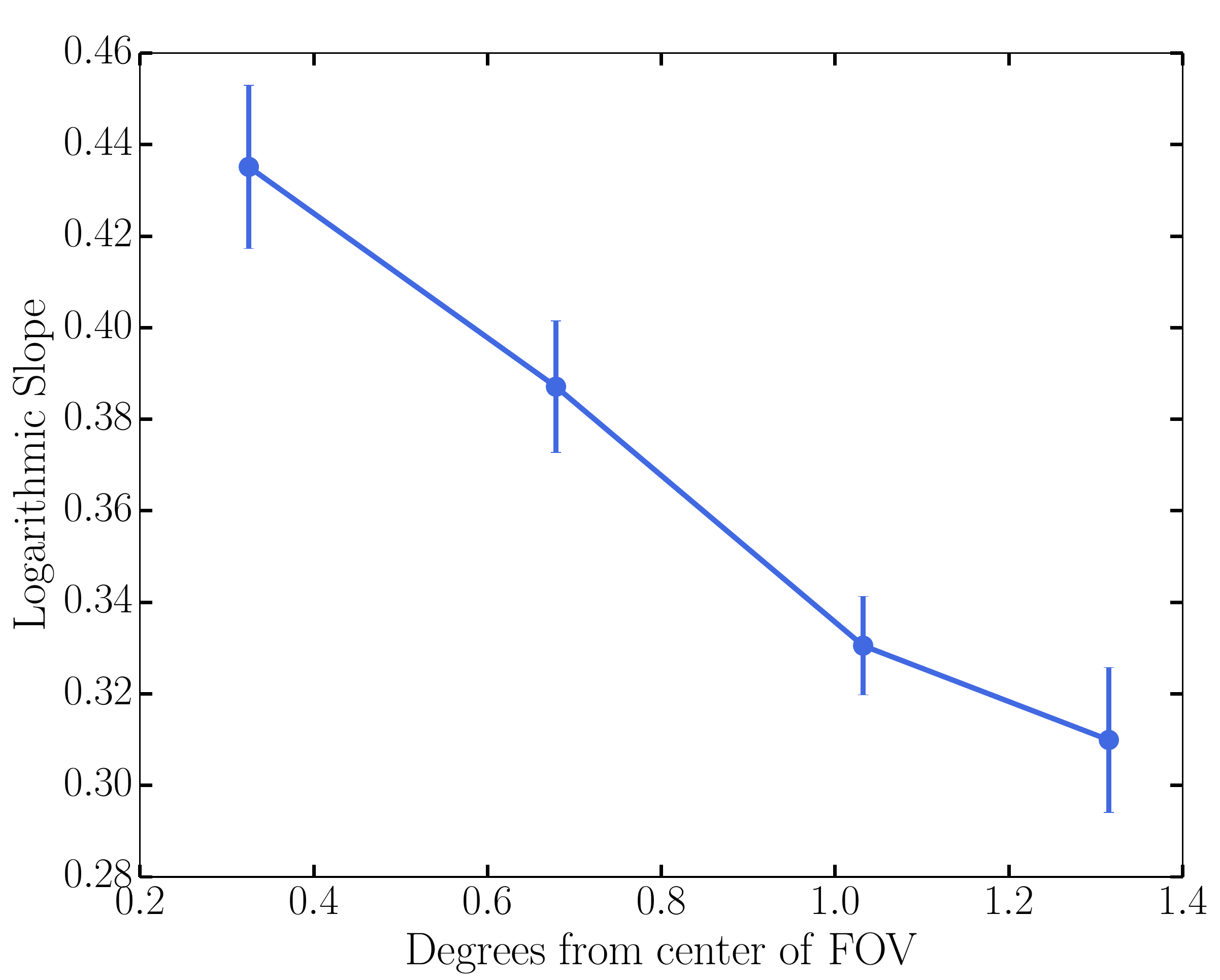}
\caption{The measured power law exponent of the size vs wavelength relation, i.e. $b$, in the $r$-band as a function of LSST focal plane position in PhoSim.  }
\label{fig:slope_v_posn_lsst}
\end{figure}

	 By simulating monochromatic sources on multiple chips at different positions in the focal plane, we find that the chromaticity is less steep further from the center of the FOV (Figure \ref{fig:slope_v_posn_lsst}). The instrumental PSF, which is mostly achromatic, is larger further from the center of the focal plane and dilutes the chromaticity from the atmosphere. The changes in Figure \ref{fig:slope_v_posn_lsst} are larger than the accuracy to which the slope needs to be measured. Thus PSF chromaticity will have to be measured as a function of focal plane position. To have $\Delta b \leq 0.02$, the FOV would have to be broken into roughly 0.3$^{\circ}$ wide annuli in the PSF analysis. Assuming 1 star arcmin$^{-2}$, this would mean $\sim$ 4000 stars in an annulus about halfway out in the focal plane. This number of stars is small enough that it will only be possible to determine the log slope to $\Delta b\sim0.03$ in $i$, allowing one to correct most, but not all, of the chromatic bias. In $r$-band it will only be possible to determine the log slope to $\Delta b\sim0.07$ if the PSF modeling errors are included. \par
	 
	 However, if a physical model of the optics can be constructed such that the instrumental PSF and chromaticity can be modeled throughout the focal plane, then all the stars in a visit can be used to infer the atmospheric chromaticity. This would allow for a very accurate determination of the atmospheric parameters. Such a model is planned as part of the PSF modeling in LSST. Additionally, while there is some evidence that the atmospheric outer scale length changes with time within a night \citep{linfield01}, it likely does not change significantly between exposures and stars across a few exposures can be used to constrain the atmospheric parameters.

\begin{figure*}
\centering
\includegraphics[scale=0.5]{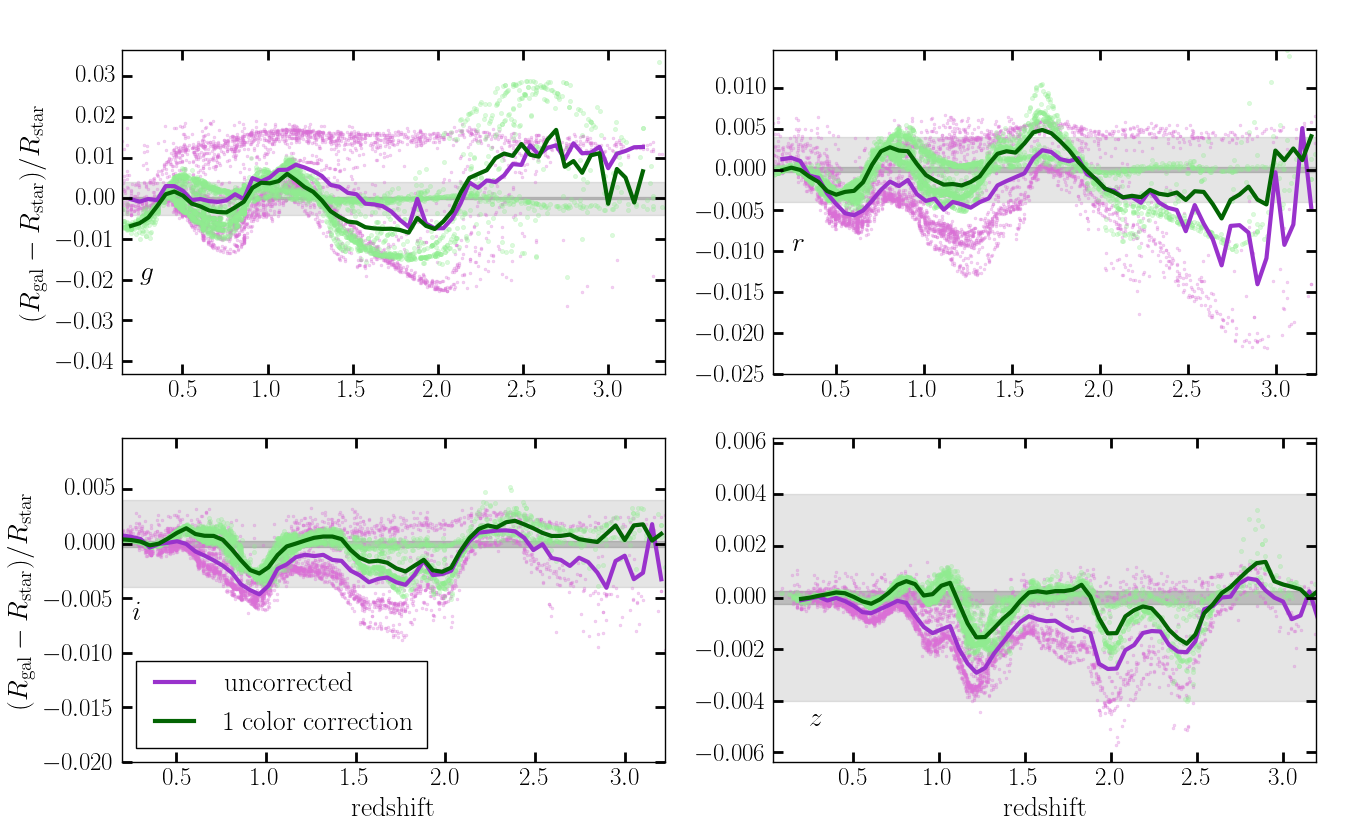}
\caption{The relative difference between the PSF size for a series of galactic SEDs and that of a K5V star SED as a function of galaxy redshift. The light gray band shows the HSC SSP first-year WL error budget of 0.004 and the dark gray band shows that of the LSST survey of $2.7\times10^{-4}$. The lines show a running mean of the points.}
\label{fig:gal_psf_err}
\end{figure*}

\section{Discussion and Summary} \label{sec:summary}
In this paper, we explored the wavelength dependence of PSFs present in a modern, optical imaging survey.  We find that redder sources have significantly smaller PSFs than blue sources in the HSC SSP survey data. This effect is at the $\sim$1 per cent level for the $g$, $r$, and $i$ bands and about an order of magnitude smaller for $z$ and $y$. We infer that the monochromatic PSF size follows a power law with wavelength of the form $\propto \lambda^{-0.2}$ to $\propto \lambda^{-0.5}$ depending on the exposure. We argue that this is consistent with the expectations of a turbulent atmosphere whose turbulent power spectrum saturates at an outer scale length of $\sim$10m-100m. We find some evidence that the level of wavelength dependence is reduced in the $g$ and $r$ bands when the atmospheric seeing is very good, because the optical/instrumental PSF in these bands is roughly achromatic. The opposite effect occurs in the $z$ band because charge diffusion in the CCD, which dominates in good seeing, has an even steeper wavelength dependence than the atmosphere. The $i$-band is intermediate; the optical/instrumental PSF has roughly the same level of chromaticity as the atmospheric PSF. The analysis of $y$-band is complicated by the fact that silicon is practically transparent to photons on the redward edge of $y$, causing photons to be absorbed after they bounce off the bottom of the silicon. This leads to large spikes present in the HSC-$y$ images. \par
Since most of the PSF chromaticity comes from the atmosphere, upcoming ground-based projects like the LSST will suffer from the same effects. \cite{meyers15} showed that a PSF wavelength dependence of $\lambda^{-0.2}$, if unmodeled, would interfere with weak lensing science since galaxies will have smaller effective PSFs than the bulk of the stars used to model the PSF. If left uncorrected, the shear will be overestimated and cosmological parameters will be biased. In Figure \ref{fig:gal_psf_err}, we show the difference between the PSF size for a variety of galactic SEDs relative to that of a K5V star assuming a monochromatic wavelength dependence of  $\lambda^{-0.45}$, typical of the worst cases we found in $r$ and $i$ in the HSC data. Since a K5V star has an intermediate color, it should be roughly representative of the PSF modeling (which is based on stars of a range of color) when no chromatic effects are included. The galactic SEDs come from the LSST CatSim project. CatSim starts with the simulated galaxy catalog of \citet{delucia2006} and attaches an SED to each galaxy based on the stellar population models of \citet{bruzual2003}. We plot 4000 galaxies with $i<25$. \citet{meyers15} give the LSST PSF size modeling requirement as $|\langle \Delta R^2 / R^2\rangle | < 5.5\times10^{-4}$, corresponding to $|\langle \Delta R / R\rangle | < 2.7\times10^{-4}$, the quantity we constrain in this paper. We also show the first year HSC WL science requirement of $|\langle \Delta R / R\rangle | \lesssim 0.004$ described in \citet{mandelbaum2017}. The average PSF error size is clearly well above the error budget for LSST and will have to be corrected for in the PSF modeling. We also show a rough correction algorithm using a single color. 2000 separate galactic SEDs drawn from the same distribution are used to derive a linear relation between the PSF size error and the SED color, assuming that the PSF size-wavelength relation is incorrect by $\Delta b \sim 0.02$ (in other words, we use size $\propto \lambda^{-0.43}$ to account for errors in measuring this relation). For the $g$ band we use $g-r$ color, for $r$ and $i$ we use $r-i$, and for $z$ we use $i-z$. This relation is used to scale the band-averaged PSF used for galaxies based on their color so that they match the fiducial stellar PSF. As can be seen in Figure \ref{fig:gal_psf_err}, this correction removes a redshift independent offset in the $r$, $i$, and $z$ bands but leaves significant redshift-dependent structure. \par

This quick demonstration shows that a more complicated algorithm that makes use of photometry in all five or six bands available is required, like the algorithms described in \citet{meyers15} and \citet{eriksen17}. These previous works showed that, if the wavelength dependence of the PSF is known to $\Delta b\sim0.02$, the chromatic bias can be corrected for to the requirements of the LSST and \textit{Euclid}. In this work we have shown that the wavelength dependence varies significantly with time due to the atmosphere but can be measured very accurately from the stars in the image frames. This work also highlights the need for a physical model of the LSST optical PSF so that all the stars across the focal plane can be used to constrain the atmospheric parameters. Such a model will also be necessary when using the measured wavelength dependence to generate `per-object' PSFs given a source SED but that is beyond the scope of the current work.\par

\section*{Acknowledgements}
We thank Rachel Mandelbaum for useful comments on this manuscript and Jim Gunn for enlightening discussions. \par
The Hyper Suprime-Cam (HSC) collaboration includes the astronomical communities of Japan and Taiwan, and Princeton University.  The HSC instrumentation and software were developed by the National Astronomical Observatory of Japan (NAOJ), the Kavli Institute for the Physics and Mathematics of the Universe (Kavli IPMU), the University of Tokyo, the High Energy Accelerator Research Organization (KEK), the Academia Sinica Institute for Astronomy and Astrophysics in Taiwan (ASIAA), and Princeton University.  Funding was contributed by the FIRST program from Japanese Cabinet Office, the Ministry of Education, Culture, Sports, Science and Technology (MEXT), the Japan Society for the Promotion of Science (JSPS),  Japan Science and Technology Agency  (JST),  the Toray Science  Foundation, NAOJ, Kavli IPMU, KEK, ASIAA,  and Princeton University.

The Pan-STARRS1 Surveys (PS1) \citep{magnier2013, tonry2012, schlafly2012} have been made possible through contributions of the Institute for Astronomy, the University of Hawaii, the Pan-STARRS Project Office, the Max-Planck Society and its participating institutes, the Max Planck Institute for Astronomy, Heidelberg and the Max Planck Institute for Extraterrestrial Physics, Garching, The Johns Hopkins University, Durham University, the University of Edinburgh, Queen's University Belfast, the Harvard-Smithsonian Center for Astrophysics, the Las Cumbres Observatory Global Telescope Network Incorporated, the National Central University of Taiwan, the Space Telescope Science Institute, the National Aeronautics and Space Administration under Grant No. NNX08AR22G issued through the Planetary Science Division of the NASA Science Mission Directorate, the National Science Foundation under Grant No. AST-1238877, the University of Maryland, and Eotvos Lorand University (ELTE).

This paper makes use of software developed for the Large Synoptic Survey Telescope. We thank the LSST Project for making their code available as free software at http://dm.lsst.org.

Based in part on data collected at the Subaru Telescope and retrieved from the HSC data archive system, which is operated by the Subaru Telescope and Astronomy Data Center at National Astronomical Observatory of Japan.

\bibliographystyle{mnras}
\bibliography{summary.bib}

\appendix

\section{Atmospheric Chromaticity with an Outer Scale}\label{app:chromo}
\subsection{Long Exposure PSF Profile}
We present a brief review of atmospheric turbulence theory in the atmosphere as it applies to observed PSFs. We include results from \citet{roddier81}, \citet{book16}, and \citet{lena12}. From the Fraunhofer diffraction equation, the observed PSF (ignoring non-diffraction-limited optical effects) will be the Fourier transform of the complex amplitude of the electric field at the pupil plane. We are interested in the PSF of long exposure images ($\gtrsim$30 sec, which is the exposure time for the LSST survey) and so we will consider a time average over the atmospheric turbulence. Assuming that the processes are ergodic, this is equivalent to a spatial average. Therefore, to predict the intensity profiles in the observed image, we need to know the spatial coherence function of the complex amplitude of the electric field of a wave, $\Psi(\vec{x})$, after it goes through a thin turbulent layer. We will later integrate over all layers. The coherence function is defined as:
\beq
B_{\Psi}(\vec{\zeta}) \equiv \langle \Psi(\vec{x}) \Psi^*(\vec{x}+\vec{\zeta}) \rangle, 
\label{eq:fieldB}
\eeq
where $\vec{x}$ is a (2-D) position on the pupil plane. For simplicity, we assume a monochromatic plane wave. Our goal is to connect this function to the power spectrum of index of refraction variations that are causing the phase shifts and thus the blurring. We will assume that the layer is thin enough so that it only introduces a phase to the field. This means we can write:
\beq
\Psi = e^{i\phi},
\eeq
since we can write $\Psi_{\infty} =1$ for the wave above the atmosphere, and the coherence function becomes
\beq
B_{\Psi}(\vec{\zeta}) = \langle e^{i(\phi(\vec{x}) - \phi(\vec{x}+\vec{\zeta}))} \rangle.
\eeq
The phase, $\phi$, should be the sum of a large number of independent perturbations (as long as the thickness of the layer is greater than the outer scale of turbulence) and therefore should be Gaussian randomly distributed (due to the central limit theorem). We can then use a simple identity for the Gaussian distributed variable to write:
\beq
B_{\Psi}(\zeta) &= \mathrm{exp}\left[-\frac{1}{2}\langle|\phi(\vec{x}) - \phi(\vec{x}+\vec{\zeta})|^2\rangle\right]\\
&\equiv \mathrm{exp}\left[-\frac{1}{2}D_{\phi}(\vec{\zeta})\right]
\label{eq:waveB}
\eeq
which defines the phase structure function, $D_{\phi}$.  
The goal is now to find the phase structure function in terms of the index of refraction variations which are causing the phase perturbations. We start with the coherence function of the phase, $\phi$:
\beq
B_{\phi}(\vec{\zeta}) = \langle \phi(\vec{x})\phi(\vec{x}+\vec{\zeta}) \rangle
\label{eq:phaseB}
\eeq
From this definition, it is easy to show that 
\beq
D_\phi(\vec{\zeta}) = 2\left[B_\phi(0) - B_\phi(\vec{\zeta})\right]
\label{eq:def_struct}
\eeq
The phase perturbation that a layer of thickness $\delta h$ causes is simply:
\beq
\phi(\vec{x}) = k \int_{h}^{h+\delta h} n(\vec{x}, z) dz
\eeq
where $k$ is the wavenumber, and $n$ is the index of refraction. Plugging this into equation \ref{eq:phaseB} we find:
\beq
B_{\phi}(\vec{\zeta}) &= k^2 \int_{h}^{h+\delta h}\int_{h}^{h+\delta h} dz\;dz' \langle n(\vec{x}, z)n(\vec{x}+\vec{\zeta}, z') \rangle\\
&= 2 k^2 \int_{h}^{h+\delta h}dz\int_{z}^{h+\delta h} dz'\; \langle n(\vec{x}, z)n(\vec{x}+\vec{\zeta}, z') \rangle
\label{eq:bphi1}
\eeq
where we have used the fact that the term in the angle brackets is symmetric to exchanging $z$ and $z'$. Defining $\xi \equiv z' - z$ and $B_n$ as the 3-dimensional covariance function of $n$, we can switch the order of integration and, assuming that $n$ varies slowly with $z$ \citep{hufnagel64, book16}, we find:
\beq
B_{\phi}(\vec{\zeta}) = 2k^2 \int_{0}^{\delta h}d\xi \int_{h}^{h+\delta h -\xi} \;dz B_n (\vec{\zeta}, \xi)
\eeq
We've assumed above that the thickness of the layer, $\delta h$, is larger than the outer scale of turbulence, so $B_n \rightarrow 0$ for $\xi > \delta h$. We can use this to simplify the integrals. Because $B_n$ is roughly independent of $z$, we can replace the $\xi$ integral upper bound with $ \infty$ and take $B_n$ out of the integral, leaving $\delta h$. Since $B_n$ is symmetric with respect to $\xi$, we can extend the bottom bound to $-\infty$ and absorb the factor of $2$:
\beq
B_{\phi}(\vec{\zeta}) = k^2 \delta h \int_{-\infty}^{+\infty} d\xi B_n (\vec{\zeta}, \xi)
\label{eq:phaseB2}
\eeq
In general, the two-point covariance function is the Fourier transform of the power spectrum of the field:
\beq
\Phi(\vec{f}) = \int d \vec{\zeta}B(\vec{\zeta})  e^{-2\pi i \vec{f}  \cdot\vec{\zeta} }
\label{eq:fourier}
\eeq
Using equation \ref{eq:phaseB2} in \ref{eq:fourier} we can relate the 2D phase power spectrum, $\Phi_{\phi}$, to the 3D power spectrum of index of refraction fluctuations, $\Phi_n$.
\beq 
\Phi_{\phi}(\vec{f}) &= \int d \vec{\zeta} k^2 \delta h \int d\xi B_n(\vec{\zeta}, \xi) e^{-2\pi i \vec{f} \cdot \vec{\zeta}} \\
&= k^2 \delta h \Phi_n(\vec{f}, 0)
\label{eq:phiphi}
\eeq
Following \citet{conan08}, we use the \citet{vonkarman} power spectrum for the index of refraction variations:
\beq
\Phi_n(\vec{f},0) = 0.0097 C_n^2 (f^2 + f_0^2)^{-11/6}
\eeq
where $C_n$ is the index of refraction \textit{structure constant}, $f\equiv|\vec{f}|$, and $f_0\equiv1/L_0$ is the inverse of the outer scale of turbulence.  This means that the power spectrum saturates on scales larger than $L_0$. $f$ is a spatial frequency with units $[m^{-1}]$.  We define $C_n$ this way to be consistent with \citet{conan08}. \par
Finally, we use this power spectrum to calculate $D_{\phi}(\vec{\zeta})$. From Equation \ref{eq:phiphi}, the phase power spectrum is
\beq
\Phi_{\phi}(\vec{f}) = 0.0097 k^2 \delta h C_n^2 (f^2 + f_0^2)^{-11/6},
\label{powerspecPhi}
\eeq
and the phase covariance function is the inverse FT of this:
\beq
B_{\phi}(\vec{\zeta}) = 0.036 k^2 \delta h C_n^2 f_0^{-5/3} (2\pi\zeta f_0)^{5/6} K_{5/6}(2\pi f_0 \zeta),
\eeq
where $K_{\nu}(x)$ is the modified Bessel function of the second kind of order $\nu$. The numerical constants come from various gamma functions that come from integrating Equation \ref{powerspecPhi}. This result is the output from one single layer, but we can combine the effect of several atmospheric layers by integration over height:
\beq
B_{\phi}(\vec{\zeta}) &=  0.036 k^2  f_0^{-5/3} (2\pi\zeta f_0)^{5/6} K_{5/6}(2\pi f_0 \zeta) \int dz C_n^2(z) \\
&=   0.036 k^2  f_0^{-5/3} (2\pi\zeta f_0)^{5/6} K_{5/6}(2\pi f_0 \zeta) \sigma^2
\eeq 

where $\sigma^2 \equiv \int dz C_n^2(z)$, with units of $[m^{-1/3}]$. This quantity is related to Fried's parameter as $r_0 = 0.1846(\lambda/\sigma)^{6/5}$. Also note that the above step implicitly assumes that the outer scale length does not change with altitude.

We can now use this in Equation \ref{eq:def_struct} to write:
\beq
D_{\phi}(\vec{\zeta}) &= 2\bigg[0.036 f_0^{-5/3} \frac{\Gamma(5/6)}{2^{1/6}} k^2\sigma^2 \\
&-  0.036 k^2  f_0^{-5/3} (2\pi\zeta f_0)^{5/6} K_{5/6}(2\pi f_0 \zeta) \sigma^2\bigg] \\
&= 0.072\left(\frac{2\pi\sigma}{\lambda}\right)^2 f_0^{-5/3} \bigg[\frac{\Gamma(5/6)}{2^{1/6}}\\
& - (2\pi f_0\zeta)^{5/6}K_{5/6}(2\pi f_0\zeta) \bigg]
\label{eq:Dphi}
\eeq
where we have used $k=2\pi/\lambda$.\par 
By taking equation \ref{eq:Dphi} to the limit of $f_0\rightarrow 0$ and using Fried's parameter, we get the phase structure function for Kolmogorov turbulence:
\beq
D_{\phi}^{\mathrm{Kolm}} (\zeta) = 6.88 \left(\frac{\zeta}{r_0}\right)^{5/3}
\label{eq:kolm_struc}
\eeq
With (either) structure function in hand, we can calculate the wavefront's covariance function via equation \ref{eq:waveB}. Note that we calculated equation \ref{eq:waveB} for the wave right after it leaves the layer. It is shown in \citet{roddier81} and \citet{lena12} that this coherence function does not change due to Fresnel diffraction through the atmosphere from this thin layer to the ground layer (it only picks up an overall phase) and so the coherence function at the telescope's pupil plane is:
\beq
B_{\Psi, 0}(\zeta) = B_{\Psi}(\zeta) = \mathrm{exp}\left[-\frac{1}{2}D_{\phi}(\vec{\zeta})\right]
\eeq
We can now simply write out the expected image in the image plane, assuming isotropy of turbulence and ignoring the modulus transfer function of the telescope \citep{book16}, as:
\beq
\langle I(\theta)\rangle = \int_{\rm aperture} B_{\Psi}(\zeta) \zeta J_0\left(\frac{2\pi\zeta\theta}{\lambda}\right) \; d\zeta
\label{eq:psf_gen}
\eeq
We use Equation \ref{eq:psf_gen} to generate the atmospheric PSFs in Sections \ref{sec:hscdata} and \ref{sec:phosim}.

\subsection{Wavelength Dependence}
We mention in the introduction that a finite outer scale length steepens the wavelength dependence of this PSF. To see the wavelength dependence of the PSF, we can consider a structure function of the general form
\beq
D_{\phi} (\zeta) = 2\beta \zeta^{\alpha},
\label{eq:Dphi_approx}
\eeq 

Using eq. \ref{eq:psf_gen} and the fact that the integral loses support for large $\zeta$, it is possible to show that the FWHM of this PSF is roughly given by
\beq
\theta_{\rm FWHM} \approx \frac{\lambda}{\pi} \sqrt{\frac{4\Gamma\left(\frac{2+\alpha}{\alpha}\right)}{\Gamma\left(\frac{4+\alpha}{\alpha}\right)}} \beta^{1/\alpha},
\label{eq:fwhm}
\eeq

If we consider the structure function for Kolmogorov turbulence (equation \ref{eq:kolm_struc}), $\beta\sim\lambda^{-2}$ and $\alpha=5/3$ so
\beq
\theta_{\rm FWHM} \sim \lambda^{-1/5}
\label{eq:fwhm-lam}
\eeq
which is the familiar result. A more accurate numerical calculation gives the FWHM in Kolmogorov turbulence as \citep{tokovinin02}:
\beq
\theta_{\rm FWHM,\; Kolm} = 0.98 \frac{\lambda}{r_0}
\eeq

\begin{figure}
\includegraphics[scale=0.3]{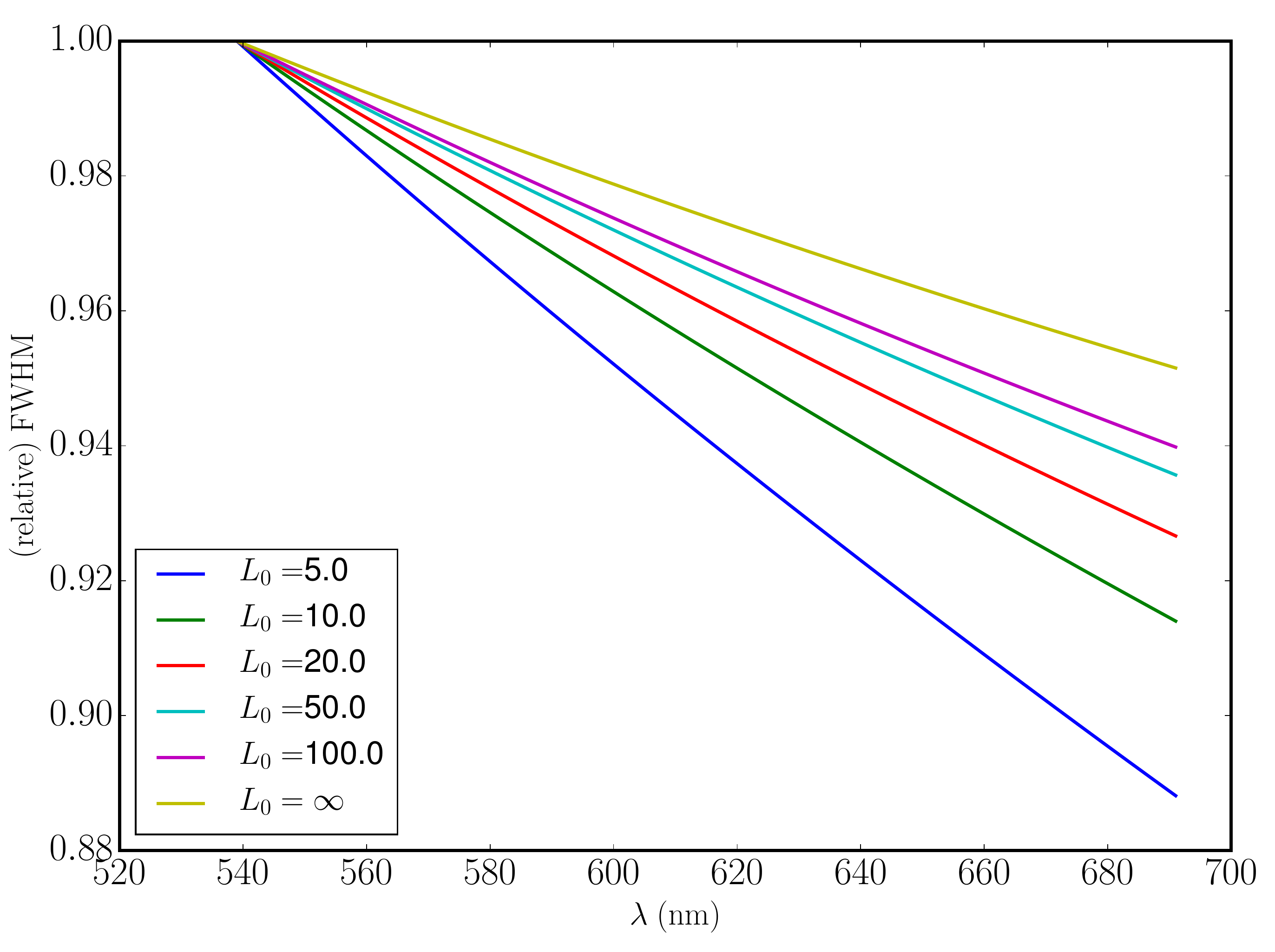}
\caption{Relative FWHM of the atmospheric PSF as a function of wavelength in the $r$-band for different outer scale lengths in meters.}
\label{fig:fw_outer}
\end{figure}

It is not easy to analytically show the wavelength dependence in the case of the more general von K\'{a}rm\'{a}n turbulence. Instead, we do the Fourier transform in Equation \ref{eq:psf_gen} numerically and measure the FWHM directly from the resulting profile. The results are shown in Figure \ref{fig:fw_outer}, where it is seen that the smaller the outer scale length, the steeper the wavelength dependence of the FWHM. Additionally, a very short outer scale length causes the wavelength dependence of the PSF to deviate from a power law and become more linear. However, for outer scale lengths $\gtrsim 10$m which is what is expected for common observatory sites, the power law approximation we made in Section \ref{sec:hscdata} is accurate.

\end{document}